\documentclass[a4paper,superscriptaddress,twocolumn,longbibliography]{revtex4-2}
\usepackage{amsmath} 
\usepackage[T1]{fontenc} 
\usepackage[utf8]{inputenc}  
\usepackage{graphicx}
\usepackage{xcolor}
\usepackage{color}
\usepackage{amssymb}
\usepackage{tikz,relsize,booktabs,tikz}
\usepackage{lmodern}
\usepackage{mathrsfs}
\usepackage{booktabs}
\usepackage{epsfig}
\newcommand{\br}{\mathbf{r}}
\newcommand{\bR}{\mathbf{R}}
\newcommand{\bk}{\mathbf{k}}
\newcommand{\bK}{\mathbf{K}}

\newcommand{\bT}{\mathbf{T}}
\newcommand{\bM}{\mathbf{M}}
\newcommand{\be}{\mathbf{e}}
\newcommand{\bn}{\mathbf{n}}
\newcommand{\bmu}{\boldsymbol{\mu}}
\newcommand{\bnu}{\boldsymbol{\nu}}

\begin{document}

\title{Wigner localization in two and three dimensions: an \emph{ab initio} approach}
\newcommand{\lcpq}{Laboratoire de Chimie et Physique Quantiques, CNRS, Universit\'e Toulouse~III (UPS), 118 Route de Narbonne, F-31062 Toulouse, France}
\newcommand{\cemes}{CEMES/CNRS, 29 rue J. Marvig, 31055 Toulouse, France}
\newcommand{\etsf}{European Theoretical Spectroscopy Facility (ETSF)}
\author{Miguel Escobar Azor}
\affiliation{\lcpq}
\affiliation{\etsf}
\author{Estefania Alves}
\affiliation{\cemes}
\author{Stefano Evangelisti}
\affiliation{\lcpq}
\author{J.~Arjan Berger}
\affiliation{\lcpq}
\affiliation{\etsf}
\email{arjan.berger@irsamc.ups-tlse.fr}

\begin{abstract}
In this work we investigate the Wigner localization of two interacting electrons at very low density in two and three dimensions using the exact diagonalization of the many-body Hamiltonian. We use our recently developed method based on Clifford periodic boundary conditions with a renormalized distance in the Coulomb potential. To accurately represent the electronic wave function we use a regular distribution in space of gaussian-type orbitals and we take advantage of the translational symmetry of the system to efficiently calculate the electronic wave function. We are thus able to accurately describe the wave function up to very low density.
We validate our approach by comparing our results to a semi-classical model that becomes exact in the low-density limit.
With our approach we are able to observe the Wigner localization without ambiguity.
\end{abstract}
\date{\today}

\maketitle
\section{Introduction}
Almost a century ago Wigner predicted that a system solely consisting of interacting electrons in a neutralizing uniform background would, at sufficiently low density, form a crystalline structure with the electrons localized at lattice sites~\cite{Wigner}. His argument can be understood by considering the dependence of the kinetic and repulsive energies on the Wigner-Seitz radius $r_s$ which is the radius of a sphere that contains, on average, one electron or, equivalently, half the average distance between nearest-neighbor electrons.
While the kinetic energy scales as $r_s^{-2}$ the repulsive energy scales as $r_s^{-1}$. As a consequence, in the low-density limit (large $r_s$) the Hamiltonian is dominated by the repulsive energy leading the electrons to localize in space. When many electrons are present the electrons will localize at crystallographic sites forming a so-called Wigner crystal. Both one and two-dimensional Wigner crystals have been observed experimentally~\cite{Grimes_1979,Shapir_2019}. More generally speaking, whenever electrons localize due to the electron-electron repulsion being dominant with respect to the kinetic energy one speaks of Wigner localization. For few electron systems one also speaks of Wigner molecules~\cite{Egger,Cioslowski_2006,Ellenberger_2006,Yannouleas_2007,Cioslowski_2017,Cioslowski_2017_JCP, Diaz-Marquez_2018,Escobar_2019} which have also been observed experimentally~\cite{Pecker}.

The main goal of this work is to study Wigner localization from an \emph{ab initio} perspective in two and three dimensions. This will allow us to better understand this phenomenon and may lead to predictions of the properties of Wigner crystals and Wigner molecules in the future. However, the inherent strong correlation in these Wigner molecules precludes the use of many standard condensed-matter theories.
In particular, Kohn-Sham density-functional theory (DFT)~\cite{HohenbergKohn,KohnSham} using standard exchange-correlation functionals fails to describe strong correlation. We note that more accurate functionals for describing strong correlation can be obtained from strictly correlated electrons DFT~\cite{Seidl_1999,Seidl_1999_2} and Wigner localization has been observed with this approach~\cite{MaletPRL,MaletPRB,Mendl_2014}. In this work we will use exact diagonalization of the many-body Hamiltonian to ensure that all correlation effects are included.
We will focus here on two-electron systems since this is sufficient to demonstrate the concept of Wigner localization.

In a recent work we used exact diagonalization to study Wigner localization in a one-dimensional system of two electrons that are confined to a ring~\cite{Escobar_2019}. In this work we generalize our approach to study Wigner localization at low densities in systems of two and three dimensions.
To study the Wigner localization in a $d$-dimensional system ($d=1,2,3$), we can represent our system in different ways. We could confine the electrons to a finite $d$-dimensional system with a positive background, but border effects would influence the results ~\cite{Diaz-Marquez_2018}.
One way of avoiding border effects is to confine the electrons to a $d$-dimensional closed space such as a $d$-torus but the numerical implementation of the geometry of a $d$-torus is cumbersome for $d>1$.
Therefore, instead, we apply periodic boundary conditions (PBC). More precisely, we define a regular $d$-dimensional supercell and then modify its topology into a toroidial topology by joining opposite sides of the cell without deformation~\cite{Tavernier_2020}.
This procedure yields a supercell that has the topology of a $d$-Clifford torus which is a flat, closed $d$-dimensional real Euclidean space embedded in a complex $d$-dimensional Euclidean space. The main difference between a torus and a Clifford torus is their curvature. While the gaussian curvature of an ordinary torus has opposite signs on different sides of the surface, the gaussian curvature of a Clifford torus vanishes at every point of the surface.
%

The paper is organized as follows. In section \ref{theory} we give the theoretical details of our approach while
in section \ref{compdet} we discuss the computational details. 
We discuss the results of our study in section \ref{results} and in section \ref{conclusions} we draw the conclusions from our work.
Throughout this work we use Hartree atomic units $\hbar=e=m_e=a_0=1$.
\section{Theory}
\label{theory}
In this section we will describe the main aspects of the theory of our approach.
\subsection{Clifford boundary conditions}
We will study a system consisting of 2 interacting
electrons confined to a $d$-dimensional Clifford torus (or flat torus).

The Hamiltonian of the system is
\begin{equation} \label{hamiltonian}
   \hat{H}=-\frac{1}{2}\nabla^2_1-
   \frac{1}{2}\nabla^2_2+
   \frac{1}{r_{12}},
\end{equation}
where the first two terms on the right-hand side correspond to the kinetic energy of electron $1$ and electron $2$,
respectively, and the last term is the 
Coulomb potential in which $r_{12}$ 
is the distance between the two electrons.
As mentioned in the Introduction, from a numerical point of view, it is convenient to apply PBC by defining a supercell with the topology of a Clifford torus
because its surface is locally flat everywhere. 
Therefore, the Laplacian, which is a local operator, is given by the usual expression $\nabla^2 = \sum_{i=1}^d \partial_i^2$
with $i$ a cartesian coordinate. We refer the reader to Ref.~[\onlinecite{Schwartz_2011}] for more details on the mathematics of Clifford tori.
%
Due to the topology of a Clifford torus there are two definitions of the distance $r_{12}$ that emerge naturally.
The first one is the geodesic distance
defined as the length of the shortest path between two points on the surface of the torus, i.e.,
\begin{equation}
     r^{geo}_{12} = \sqrt{\sum_{i=1}^d r^2_{{12}}(i)} 
  \label{geodesic_distance},
\end{equation}
where
\begin{align}
\label{Eqn:r12}
   &  r_{12} (i) = \left\{
	       \begin{array}{ll}
		 |{r_1}(i)-{r_2}(i)|    
		 & \mathrm{if\ } |{r_1}(i)-{r_2}(i)| < \frac{L}{2} \\
		 \\
		 L - |{r_1}(i)-{r_2}(i)| & \mathrm{if\ } |{r_1}(i)-{r_2}(i)| > \frac{L}{2} \\
        \end{array}
	     \right.,
 \end{align}
in which $L$ is the length of an edge of the Clifford supercell (CSC) and $i$ is a Cartesian component. For convenience we have chosen the length of all the edges to be equal.
The disadvantage of this definition is that the derivative of $r_{12}^{geo}$ with respect to $i$ is discontinuous in those points where $r_{12}(i)=L/2$\cite{Tavernier_2021}.
Therefore, also the forces, which are related to the gradient of the Coulomb potential, are discontinuous in those points. This is an unphysical result.





The second natural definition of the distance $r_{12}$ is the Euclidean (or straight-line) distance, defined as the length of the shortest possible path between two points in the embedding space of the Clifford torus,
%
%
\begin{equation} 
    \label{euclidian_distance}
    r^{euc}_{12} =\frac{L}{\pi} \sqrt{ \sum_{i=1}^d \sin^2  \bigg(r_{12}(i)\frac{\pi}{L} \bigg)},
\end{equation} 
where $r_{12}(i)$ was defined in Eq.~\eqref{Eqn:r12}. 

In the following we will use the Euclidean distance in the definition of the Coulomb potential since it is smooth and continuously differentiable. We have previously successfully applied this strategy 
in the calculation of Madelung constants~\cite{Tavernier_2020,Tavernier_2021} and in the calculation of the ground-state energies of Wigner crystals~\cite{Alves_2021}. Finally, we note that the Euclidean distance defined above is related to a position operator that we have recently proposed that is compatible with PBC~\cite{Valenca_2019}.
\subsection{Exact diagonalization}
We solve the time-independent Schr\"odinger equation involving the Hamiltonian in Eq.~\eqref{hamiltonian} by employing an exact-diagonalization approach. Therefore, we project the Hamiltonian onto the basis of 2-electron Slater determinants and diagonalize the resulting Hamiltonian matrix to obtain the wave functions and eigenenergies.
In the remainder of this subsection we discuss how we build the 2-electron Slater determinants from a symmetry-adapted basis set. This allows us to perform the calculations in a numerically efficient manner such that we can accurately describe the Wigner localization by employing a large number of basis functions.
\subsubsection{The basis set}
Since we want to describe a localization effect of the electrons it is convenient to use a localized basis set instead of a delocalized one, e.g., plane waves.
In particular, the localized gaussian-type orbitals (GTO's) are convenient in practice because of the gaussian product theorem~\cite{Szabo}.
These are the main reasons why we choose gaussian functions as a primitive basis set to study Wigner localization.
We note that, strictly speaking, a gaussian function is defined everywhere in space,
while the functions that we need to describe the electrons in the CSC must be periodic 
smooth functions confined to the torus.
A gaussian function, which is not a function with a compact support, 
is not a continuous and differentiable function that is defined everywhere on the torus.
However, if the width of the gaussian is much smaller than the size of the torus then, 
for any practical purpose, this function can be considered as a 
$\cal C^\infty$ function having a compact support on the torus.
Therefore, by combining various gaussians centered on a regular grid on the torus, 
we are able to build periodic smooth wave functions on this manifold.



We use $m^d$ three-dimensional $s$-type GTO's that are evenly distributed on a regular $d$-dimensional grid 
with $m$ the number of gaussians placed along each side of the Clifford supercell.
Moreover, we choose all gaussians to have a common exponent. 
The normalized three-dimensional $s$-type gaussians functions are defined as
\begin{equation} 
\label{gaussian}
g_{\boldsymbol{\mu}}({\mathbf{r}- \mathbf{R}_{\boldsymbol{\mu}}})  =
\bigg(\frac{2\alpha}{\pi}\bigg)^\frac{3}{4}
e^{-\alpha |\mathbf{r}- \mathbf{R}_{\boldsymbol{\mu}}|^2},
\end{equation}
where $\alpha$ is the exponent of the gaussian and  $\mathbf{R}_{\boldsymbol{\mu}}$ is the center of the gaussian ${\boldsymbol{\mu}}$.
For convenience the gaussians are labeled with a vector ${\boldsymbol{\mu}}$ containing $d$ integers, one for each dimension of the system.
We note that we use the three-dimensional gaussians also in the case of one- and two-dimensional systems.
Therefore, the one- and two-dimensional systems studied in this work are in fact quasi-1- and quasi-2-dimensional systems.

Since the electrons are confined to the Clifford torus also the basis functions that form the electronic wave function have to be confined to the Clifford torus. As a consequence, the gaussian product rule mentioned before has to be applied using the geodesic distance in order to guarantee that resulting gaussians also lie on the Clifford torus. More precisely the distance 
$R_{\bmu\bnu}$ between two gaussians $g_{\boldsymbol{\mu}}$ and $g_{\boldsymbol{\nu}}$ is defined as
\begin{equation}
    R_{\boldsymbol{\mu}\boldsymbol{\nu}} = \sqrt{\sum_{i=1}^d R^2_{{\boldsymbol{\mu}\boldsymbol{\nu}}}(i)},
\end{equation}
where
\begin{align} 
   &  R_{\boldsymbol{\mu}\boldsymbol{\nu}} (i) = \left\{
	       \begin{array}{ll}
		 |{R_{\boldsymbol{\mu}}}(i)-{R_{\boldsymbol{\nu}}}(i)|    
		 & \mathrm{if\ } |{R_{\boldsymbol{\mu}}}(i)-{R_{\boldsymbol{\nu}}}(i)| < \frac{L}{2} \\
		 \\
		 L - |{R_{\boldsymbol{\mu}}}(i)-{R_{\boldsymbol{\nu}}}(i)| & \mathrm{if\ } |{R_{\boldsymbol{\mu}}}(i)-{R_{\boldsymbol{\nu}}}(i)| > \frac{L}{2} \\
        \end{array}
	     \right..
\end{align}
It is clear that the accuracy of the numerical results will depend on the overlap between two nearest-neighbor gaussians. If the overlap is too small the basis set will not be able to correctly describe the electronic wave function while if it is too large, i.e., close to 1, a quasi-linear dependence of the basis functions might lead to numerical problems. 
Since all gaussians have the same exponent $\alpha$ the nearest-neighbor overlap is simply given by $e^{-\xi/2}$ in which $\xi=\alpha\delta^2$ with $\delta$ the distance between two nearest-neighbor gaussians.
We have analyzed this issue in much detail in previous works and we have derived optimal values of $\xi$ for systems of different dimensions~\cite{Brooke_2018,Escobar_2019}.


%
\subsubsection{Symmetry-adapted orbitals}
As a consequence of the periodicity 
of the Clifford supercell, the system is translationally invariant. 
Therefore, the translation operator, $\hat{T}_R$, commutes with the Hamiltonian, $\hat{H}$, and the eigenstates of $\hat{H}$ can be chosen to be equal to the eigenstates of $\hat{T}_R$.
It is hence convenient to construct symmetry-adapted orbitals (SAO) from a linear combination of GTO's that satisfy the translational invariance~\cite{Angeli_2021}. The (unnormalized) SAO are defined as
\begin{equation}
\label{Eqn:SAO}
 \phi_{\mathbf{k}}(\mathbf{r})  = \frac{1}{m^{d/2}}
  \sum_{{\boldsymbol{\mu}}}
 e^{ i \frac{2\pi}{m} \mathbf{k} \cdot \boldsymbol{\mu}} 
 g_{\boldsymbol{\mu}}
 ( \mathbf{r}- \mathbf{R}_{\boldsymbol{\mu}}),
\end{equation}
where $\mathbf{k}=(k_1,\cdots,k_d)^T$ with $k_i = 0, \cdots, m-1$.
We refer the reader to Appendix \ref{App:SAO} for a proof of the fact that $\hat{T}_R\phi_{\mathbf{k}}(\mathbf{r}) = \phi_{\mathbf{k}}(\mathbf{r})$.

The one- and two-electron integrals in the symmetry-adapted basis are given by, respectively,

\begin{align}
& T_{\mathbf{k}, \mathbf{k}'} =
 \delta_{\mathbf{k}, \mathbf{k'}} \mathcal{S}_{\bk}^{-1}
 \sum_{\boldsymbol{\nu}}
 \cos \left[\frac{2\pi}{m}(\mathbf{k} \cdot \boldsymbol{\nu}) \right]
 T_{\mathbf{0},\boldsymbol{\nu}},
 \label{one-electron integral}
\\
\nonumber
 & \langle \mathbf{k}, \mathbf{k'} | \mathbf{k''}, \mathbf{k'''}  \rangle  =
 \frac{1}{m^d} \delta_{\mathbf{k}+\mathbf{k'} -\mathbf{k''} -\mathbf{k'''}} [\mathcal{S}_{\bk}\mathcal{S}_{\bk'}\mathcal{S}_{\bk''}\mathcal{S}_{\bk'''}]^{-1/2}
 \\ & \times
 \sum_{\boldsymbol{\nu \rho \sigma}}
  \cos\left[\frac{2\pi}{m}(\mathbf{k'} \cdot \boldsymbol{\nu} - \mathbf{k''} \cdot \boldsymbol{\rho} - 
 \mathbf{k'''} \cdot \boldsymbol{\sigma}) \right] 
  \langle \mathbf{0}, \boldsymbol{\nu} | \boldsymbol{\rho} , \boldsymbol{\sigma} \rangle,
  \label{two-electron integral}
\end{align}
in which $\mathcal{S}_{\bk}=\langle\phi_{\bk}|\phi_{\bk}\rangle$ and where we defined
\begin{align}
\label{kin_AO}
T_{{\boldsymbol{\mu},\boldsymbol{\nu}}} & = - \frac{1}{2} \int{ 
    g_{\boldsymbol{\mu}} (\mathbf{r}_1-\mathbf{R}_{\boldsymbol{\mu}})
    \nabla^2_1 g_{\boldsymbol{\nu}}
    (\mathbf{r}_1-\mathbf{R}_{\boldsymbol{\nu}}) d\mathbf{r}_1},
\\
\nonumber
\langle \boldsymbol{\mu},\boldsymbol{\nu}|\boldsymbol{\rho},\boldsymbol{\sigma} \rangle  &= 
     \iint 
    g_{\boldsymbol{\mu}} (\mathbf{r}_1-\mathbf{R}_{\boldsymbol{\mu}})
    g_{\boldsymbol{\nu}} (\mathbf{r}_2-\mathbf{R}_{\boldsymbol{\nu}}) \times 
\\ &
    \frac{1}{r^{euc}_{12}}
       g_{\boldsymbol{\rho}} (\mathbf{r}_1-\mathbf{R}_{\boldsymbol{\rho}})
    g_{\boldsymbol{\sigma}} (\mathbf{r}_2-\mathbf{R}_{\boldsymbol{\sigma}}) 
     d\mathbf{r}_1 d\mathbf{r}_2.
\label{two_e_AO}
\end{align}
%
%
The details of the derivation of the one- and two-body integrals can be found in Appendix \ref{App:integrals}.
With the above integrals we can now express the Hamiltonian in the basis of the Slater determinants $\Phi_{\mathbf{k,k'}}=|\phi_{\mathbf{k}} \bar{\phi}_{\mathbf{k'}} \rangle$.
Owing to the Slater-Condon rules these matrix elements can be split in three different cases~\cite{Szabo}.
\\
\emph{Case 1}: The two Slater determinants in the matrix element of the Hamiltonian are identical
%
\begin{equation}
 \langle  \Phi_{\mathbf{k,k'}} |\hat{H}| \Phi_{\mathbf{k,k'}} \rangle = T_{\mathbf{k},\mathbf{k}} +
 T_{\mathbf{K}-\mathbf{k},\mathbf{K}-\mathbf{k}} + \langle \mathbf{k}, \mathbf{K}-\mathbf{k} | \mathbf{k},\mathbf{K}-\mathbf{k}\rangle
\end{equation}
where we defined $\mathbf{K}=\mathbf{k}+\mathbf{k}'$.
\\
\emph{Case 2}: The two Slater determinants differ by one spinorbital, i.e., $\mathbf{k}\neq\mathbf{k}''$,
\begin{equation}
 \langle  \Phi_{\mathbf{k,k'}} |\hat{H}| \Phi_{\mathbf{k'',k'}} \rangle = 
 T_{\mathbf{k,k''}} +
 \langle \mathbf{k}, \mathbf{k'} | \mathbf{k''}, \mathbf{k'}\rangle = 0
\end{equation}
where in the last step we used Eqs. \eqref{one-electron integral} and \eqref{two-electron integral}.
Similarly we have that $\langle  \Phi_{\mathbf{k,k'}} |\hat{H}| \Phi_{\mathbf{k,k''}}\rangle = 0$ for $\mathbf{k}'\neq\mathbf{k}''$.
\\
\emph{Case 3}: The two Slater determinants differ by two spinorbitals, i.e., $\mathbf{k}\neq\mathbf{k}'\neq\mathbf{k}''\neq\mathbf{k}'''$,
\begin{equation}
 \langle  \Phi_{\mathbf{k,k'}} |\hat{H}| \Phi_{\mathbf{k'',k'''}} \rangle = 
 \langle \mathbf{k},\mathbf{k'} | \mathbf{k''}, \mathbf{k'''}\rangle = 
  \langle \mathbf{k},\mathbf{K}-\mathbf{k} | \mathbf{k}'',\mathbf{K}-\mathbf{k}''\rangle
\end{equation}
since $\mathbf{K}=\mathbf{k}+\mathbf{k}'=\mathbf{k}''+\mathbf{k}'''$ due to the Kronecker delta that appears on the right-hand side of Eq. \eqref{two-electron integral}.

From the above considerations we observe that 
two Slater determinants with different $\mathbf{K}$ do not contribute to the same wave function. 
As a consequence the resulting Hamiltonian matrix is block diagonal with a block for each $\mathbf{K}$.
We can therefore distinguish the solutions of the Schr\"odinger equation by the different values of $\mathbf{K}$ and we can write the wave functions according to
%

%
\begin{equation}
\label{psi_k}
\begin{split}
  \Psi_{\boldsymbol{K}} (\mathbf{r}_1,\mathbf{r}_2) &= \frac{1}{\sqrt{2}}\sum_{\boldsymbol{k}} C_{\boldsymbol{k},\boldsymbol{K}-{\boldsymbol{k}}} \times\\
  & \bigg [
 \phi_{\boldsymbol{k}}(\mathbf{r}_1) {\bar{\phi}}_{\boldsymbol{K}-
 {\boldsymbol{k}}}(\mathbf{r}_2) - 
 \phi_{\boldsymbol{k}}(\mathbf{r}_2) 
 {\bar{\phi}}_{\boldsymbol{K}-{\boldsymbol{k}}}(\mathbf{r}_1) \bigg ]   
\end{split}
\end{equation}
where, $C_{\mathbf{k},\boldsymbol{K}-{\mathbf{k}}}$ are coefficients that are obtained from the diagonalization of the Hamiltonian matrix.
We note that the factor $2\pi\mathbf{k}/m$ can be interpreted as a quasimomentum that is conserved. Therefore, $2\pi\mathbf{K}/m$ can be interpreted as the total quasimomentum, which is also conserved.
The eigenenergies corresponding to the various $\mathbf{K}$ blocks of the Hamiltonian 
hence only differ in the kinetic energy of the center of mass of the system.
In the following we can, therefore, focus exclusively on the block $\mathbf{K}=\mathbf{0}$ without loss of generality and we define $\Psi(\mathbf{r}_1,\mathbf{r}_2)=\Psi_{\mathbf{K}=\mathbf{0}}(\mathbf{r}_1,\mathbf{r}_2)$.
\subsubsection{Spin adaptation}
For a two-electron system, the electronic wavefunction can correspond either to a spin singlet or to a spin triplet.
In the limit of vanishing density the singlet and triplet states become degenerate.
Therefore, at very low density both states can become quasi-degenerate.
This could yield wave functions that are mixtures of the two states due to round-off errors in the numerical calculation. 
As a consequence the computed wave functions would not be eigenfunctions of $\hat{S}^2$.
To avoid such problems, we performed a spin adaptation of the wave function.

In the case of two electrons an antisymmetrized wave function can either have a symmetric spin component (singlets)  
or an antisymmetric spin component (triplets).
Therefore, the symmetric and anti-symmetric combinations ($\mathbf{K}=\mathbf{0}$) are given by, respectively,
%
 %
  \begin{align}
  \Phi_{
  \boldsymbol{\mathbf{k}},-{\boldsymbol{\mathbf{k}}}}^S &=
  \frac{1}{\sqrt{2}} \big [ |\phi_{\boldsymbol{\mathbf{k}}}(\bf{r}_1) \bar{\phi}_{-{\boldsymbol{\mathbf{k}}}}(\bf{r}_2) \rangle
  + |\bar{\phi}_{-{\boldsymbol{\mathbf{k}}}}(\bf{r}_1) \phi_{\boldsymbol{\mathbf{k}}}(\bf{r}_2) \rangle \big ]
  \\
    \Phi_{
  \boldsymbol{\mathbf{k}},-{\boldsymbol{\mathbf{k}}}}^T &=
  \frac{1}{\sqrt{2}} \big [ |\phi_{\boldsymbol{\mathbf{k}}}(\bf{r}_1) \bar{\phi}_{-{\boldsymbol{\mathbf{k}}}}(\bf{r}_2) \rangle
  - |\bar{\phi}_{-{\boldsymbol{\mathbf{k}}}}(\bf{r}_1) \phi_{\boldsymbol{\mathbf{k}}}(\bf{r}_2) \rangle \big ]
 \end{align}
The singlet and triplet wave functions are orthogonal and can, therefore, be calculated separately, thus completely
avoiding any possibility of finding mixed-state solutions.

\subsection{2-RDM}
Since we are considering a floating Wigner system the one-body density is constant in the Clifford supercell and, therefore, not appropriate to characterize the Wigner localization. We note that there also exists pinned Wigner systems in which the wave function has a broken symmetry~\cite{Drummond_2004}. For such systems the one-body density is sufficient to show the Wigner localization.
Instead, to demonstrate the Wigner localization at low electron density we calculate the diagonal elements of the two-body reduced-density matrix (2-RDM) since for a given position of one electron it expresses the probability of finding the other electron as a function of its position.
In general, the 2-RDM is defined as

\begin{equation}
\begin{split}
   & \Gamma^{(2)}(\mathbf{r}_1, \mathbf{r}_2; \mathbf{r}'_1, \mathbf{r}'_2)=
    \frac{(N-1)N}{2} \times \\
   & \int {\Psi_N^*(\mathbf{r'}_1, \mathbf{r'}_2, \mathbf{r}_3,\cdots, \mathbf{r}_N)
    \Psi_N(\mathbf{r}_1,\cdots, \mathbf{r}_N) }
    d\br_3 \cdots d\br_N 
\end{split}
\end{equation}
in which $\Psi_N$ is an $N$-body wave function. 

For two electrons in the Clifford supercell the diagonal elements of the 2-RDM, i.e., $\Gamma(\mathbf{r}_1, \mathbf{r}_2) = \Gamma^{(2)}(\mathbf{r}_1, \mathbf{r}_2; \mathbf{r}_1, \mathbf{r}_2)$, are given by
\begin{equation}
    \Gamma(\mathbf{r}_1, \mathbf{r}_2) =
   \left|\Psi(\mathbf{r}_1, \mathbf{r}_2)\right|^2
\end{equation}
where $\Psi$ is the two-electron wave function defined in Eq.~\eqref{psi_k} (for $\bK=\mathbf{0}$).
It is useful to express $\Gamma$ in the gaussian basis set according to
\begin{equation}
    \Gamma_{\boldsymbol{\mu},\boldsymbol{\nu}} =
    \iint 
    g_{\boldsymbol{\mu}}(\mathbf{r}_1)
    g_{\boldsymbol{\nu}}(\mathbf{r}_2)
    \Gamma
    (\mathbf{r}_1,
    \mathbf{r}_2)
    d{\mathbf{r}_1} d{\mathbf{r}_2},
\end{equation}
since it describes the probability of the presence of an electron inside the gaussian $g_{\boldsymbol{\nu}}$ 
when the other electron is inside the gaussian $g_{\boldsymbol{\mu}}$.
\subsection{Semi-classical model}
To validate our approach we will compare our results to those obtained with a semi-classical model that tends to the exact solution in the limit of vanishing density, i.e., in the strong-interaction limit. In this limit we can Taylor expand the Coulomb potential in the Hamiltonian of Eq.~\eqref{hamiltonian} around the equilibrium positions $\bR_1$ and $\bR_2$ of the two localized electrons in a classical Wigner system according to
\begin{equation}
\label{Hamiltonian_exp}
\hat{H} = \underbrace{-\frac12 \sum_{i=1,2} \nabla_{\br_{i}}^2 + \hat{U}_0 + \hat{U}_2}_{\hat{H}_0} +  \hat{U}_3+ \hat{U}_4
+ \cdots
\end{equation}
in which
\begin{align}
\hat{U}_0 &=\frac{1}{R_{12}} 
\\
\notag
\hat{U}_2 &=\frac12\sum_{i,j}\sum_{\alpha\beta}
\left. \partial_{i\alpha}\partial_{j\beta} \frac{1}{r_{12}}\right|_{{\mathbf{r}}_{x}={\mathbf{R}}_{x}\forall x} 
\\ & \times 
({r}_{i,\alpha} -R_{i,\alpha}) ({r}_{j,\beta} - R_{j,\beta}),
\label{Eqn:U2}
\end{align}
where the Greek letters $\alpha$ and $\beta$ denote Cartesian components.
Therefore, the Hamiltonian $\hat{H}_0$ contains both the classical Coulomb interaction 
and a harmonic correction due to the zero-point motion while $\hat{U}_n$ ($n>3$) are anharmonic corrections that contain $n$-th order derivatives of the Coulomb potential.
We note that the first-order term in the Taylor expansion, $\hat{U}_1$, vanishes because with the electrons at their equilibrium positions the energy is at a minimum.

The energy $\hat{U}_0$ is the classical energy of two electrons located at their equilibrium positions. In the strong interaction limit the two electrons will be as far apart as possible in the $d$-dimensional Clifford supercell. Therefore $R_{12}$ attains its maximum value of $L\sqrt{d}/\pi$ and $U_0$ becomes
\begin{equation}
\hat{U}_0 = \frac{\pi}{L\sqrt{d}}.
\end{equation}

The remainder of $\hat{H}_0$ represents the zero-point correction to the classical energy $U_0$ in the harmonic approximation. 
Using a normal-mode transformation one can show that it is equivalent to the Hamiltonian of a quantum harmonic oscillator with the following harmonic frequency~\cite{Alves_2021}
\begin{equation}
\label{Eqn:harm_freq}
    \omega=\sqrt{\frac{2\pi^3 }{(d L^2)^{\frac{3}{2}}}}.
\end{equation}
%
Therefore the eigenenergies $E_{\bn}$ of $\hat{H}_0$ correspond to 
%
%
\begin{equation}
\label{Eqn:model_energies}
E_{\bn} = \frac{\pi}{L\sqrt{d}} +
\sqrt{\frac{2\pi^3}{(d L^2)^{3/2}}}
\sum_{i=1}^d \left(n_i + \frac12\right),
\end{equation}
in which $\bn$ is a vector containing $d$ non-negative integers $n_i$.
The smallest eigenenergy $E_{\mathbf{0}}$ of $\hat{H}_0$ can thus be written as
\begin{equation}
E_{\mathbf{0}} = \frac{\pi}{L\sqrt{d}} + \frac{d}{2} \sqrt{\frac{2\pi^3}{(d L^2)^{3/2}}}.
\end{equation}

The anharmonic corrections to the energy can be calculated using perturbation theory. 
It can be shown that the first-order correction to the energy due to $\hat{H}_3$, vanishes.
Therefore, the lowest-order correction to the energy consists of $U_3^{2}$, i.e.,  the second-order correction to the energy due to $\hat{U}_3$ and $U_4^{1}$, i.e.,  the first-order correction to the energy due to $\hat{U}_4$. 
They are both proportional to $L^{-2}$. 
Furthermore, in the case of 2 electrons, one can show that $U_3^{2}$ vanishes.
Therefore, the lowest-order anharmonic correction for two electrons is entirely due to $U_4^{1}$.
For the ground state this correction is given by
\begin{equation}
U_4^{1}= \langle \Psi_0 |\hat{U}_4| \Psi_0 \rangle =\frac{(6-d) \pi^2}{16 L^2},
\label{Eqn:U4_pert}
\end{equation}
where the ground-state wave function $\Psi_0$ is a product of $d$ ground-state wave functions of the quantum harmonic oscillator with the frequency given in Eq.~\eqref{Eqn:harm_freq}.
%
The total ground-state energy $E_{gs}$ can thus be written as
%
%
\begin{equation}
E_{gs} = \frac{1}{\sqrt{d}}\left(\frac{\pi}{L}\right) 
+ \frac{d^{1/4}}{\sqrt{2}} \left(\frac{\pi}{L}\right)^{3/2} 
+ \frac{(6-d)}{16}\left(\frac{\pi}{L}\right)^{2}
+ \mathcal{O}(L^{-5/2}).
\end{equation}

Finally, we note that in the 1D case the general expression for $U_4^0$ including the corrections of the energies of the excited states is given by
\begin{equation}
U_4^{1}(n) = \frac{1}{16}\left(\frac52 + 10 \left(n+\frac12\right)^2\right) \left(\frac{\pi^2}{L^2}\right)
\label{Eqn:anharm_1D},
\end{equation}
where $n$ is a non-negative integer.
%
%
%
\section{Computational Details}
\label{compdet}
As mentioned before, we use a gaussian basis set in which the (normalized) gaussians are equally spaced on a regular grid. 
%
%
In previous works~\cite{Diaz-Marquez_2018,brooke2018distributed} we have shown that the overlap $S$ 
between two neighboring gaussians  is inversely proportional to the dimensionless quantity $\xi=\alpha{\delta^2}$, where $\delta$ is the distance between the centers of two nearest neighbor gaussians, according to
\begin{align} 
S(\xi)&=e^{-\xi/2} .
\end{align}
Therefore, for large $\xi$, the overlap between two nearest neighbor gaussians will be too small to accurately describe the electronic wave function.
On the other hand, for small $\xi$, the overlap between two nearest neighbor gaussians will approach unity and the basis
functions become quasi-linear dependent which could lead to numerical instabilities in the calculations.
Due to these two limiting cases, we have to choose $\xi$ such that the basis set gives us an accurate description of the electronic wave function while, at the same time, avoiding numerical instabilities due to quasi-linear dependencies of the basis functions.
In a previous work~\cite{brooke2018distributed} we discussed in detail the range of $\xi$ for which the electronic wave function is
accurately described.
In accordance with those results and after further numerical investigations we have chosen $\xi=0.8$ in this work.

If we do not impose any approximations the calculations remain numerically feasible up to about 1000 GTO's.
This means that in 2D and 3D the number of gaussians placed along each edge are about 30 and 10, respectively.
With 1000 GTO's we can accurately describe systems of the following sizes $L=10^6$ bohr in 1D, $L=10^4$ bohr in 2D and $L= 10^3$ bohr in 3D, with $L$ the length of an edge of the CSC.
Indeed, for larger values of $L$ the average width of the electron distribution becomes of the same order 
of magnitude or even smaller than the width of the gaussians, and our description breaks down.

If we want to describe larger system sizes, we must increase the number of gaussians.
In order to do this, we can apply a controlled approximation 
that is related to the fact that many of the two electron integrals given in Eq.~\eqref{two-electron integral} are so small
that they can be neglected without changing the results.
In fact, only those two-electron integrals for which the gaussian center $\boldsymbol{\rho}$ is close to the center $\bf{0}$ \emph{and} 
the gaussian center $\boldsymbol{\sigma}$ is close to the center $\boldsymbol{\nu}$ will have non-negligible contributions to the final results.
Therefore, in practice, we will only include the two-electron integrals for which $ -\lambda \leq \rho_i \leq \lambda$ \emph{and} 
$-\lambda \leq \sigma_i - \nu_i \leq \lambda$ for all $i=x,y,z$.
After numerical investigations, we have chosen $\lambda= 6$ since it leads to relative errors in the energies of only $10^{-8}$ Hartree.
This approach allows us to extend the number of basis functions in 2D and 3D to $10000$ and $8000$ GTO's, respectively, allowing us to describe systems of length $L=10^6$ bohr in 2D and $L=10^4$ bohr in 3D.
We note that an alternative approach to increase the accuracy could be to extrapolate the results obtained for finite basis sets to the complete basis set limit~\cite{Cioslowski_2021}.


Finally, as mentioned before, due to the 3D nature of the gaussian orbitals, the one- and two-dimensional 
systems are in fact quasi-1- and quasi-2-dimensional systems.
Therefore, in order to compare our results with those obtained for the pure 1D and 2D systems of the semi-classical model, 
we have to subtract the contribution to the energy due to the transversal component(s) of the 3D gaussians.
For a 1D and 2D systems, the transversal energy per electron is equal to $\alpha$ and $\frac{\alpha}{2}$ 
respectively, where $\alpha$ is the exponent of the gaussian.~\cite{brooke2018distributed}

\section{Results}
\label{results}
In this section we present the results of our approach.
First, we will validate the implementation of our method by comparing its results to those obtained within the semi-classical model.
Then we will report the 2-RDM and show that with our approach we can capture the Wigner localization.
\subsection{Validation}
\begin{figure}[h]%
    \centering
    \includegraphics[width=\columnwidth]{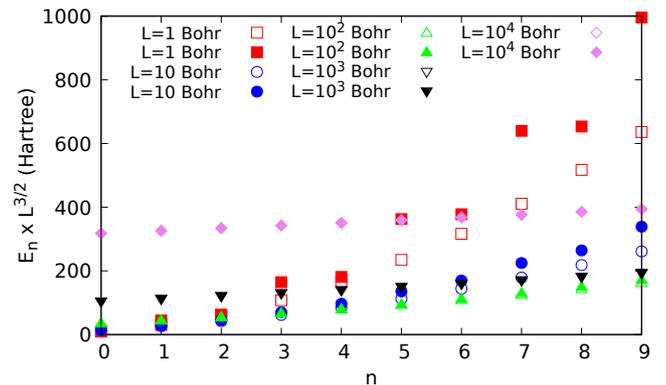}
    \caption{ Scaled 
    energies ($E_n \times L^{3/2}$) of the 10 first energy levels ($0\ge n \ge 9$) for 1-dimensional CSC of various sizes. 
    Open symbols: semi-classical model for the the low-density regime; filled symbols: exact diagonalization of the Hamiltonian. When the open symbols are not visible it means that the results obtained with the model and the exact diagonalization completely overlap.}
    \label{1D_v1}
\end{figure}

Let us first analyze the energy spectrum of 1D system of various sizes. 
In Fig.~\ref{1D_v1} we collected the 10 lowest energies for 1D systems of several sizes obtained from our \emph{ab initio} calculations and from the semi-classical model. 
As expected, there are important differences between the energy spectra obtained with both methods when $L$ is small,
since for small $L$ (high density) the semi-classical model, which is based on an asymptotic expansion around $L\rightarrow\infty$, is not appropriate.
Instead, for large $L$ the agreement between the energy spectra obtained with the model and with the \emph{ab initio} calculations is excellent.
When the system size gets larger, the difference between the numerical results and those obtained with the model gets smaller since contributions beyond the first anharmonic corrections fall off at least as $L^{-5/2}$.

Finally, we see from Fig.~\ref{1D_v1} that for a given value of $L$ the difference between the \emph{ab initio} and model energies increases with the level of the excited state $n$.
This is in agreement with the expressions given in Eqs.~\eqref{Eqn:model_energies} and \eqref{Eqn:anharm_1D} 
which suggest that a term of order $L^{-p/2}$ is proportional to $n^{p-2}$ with $p\ge 2$ an integer.
In other words, the higher the excited state becomes, the more important the higher-order anharmonic corrections that are not included in the model become.

%
\begin{figure}[h]%
    \centering
    \includegraphics[width=\columnwidth]{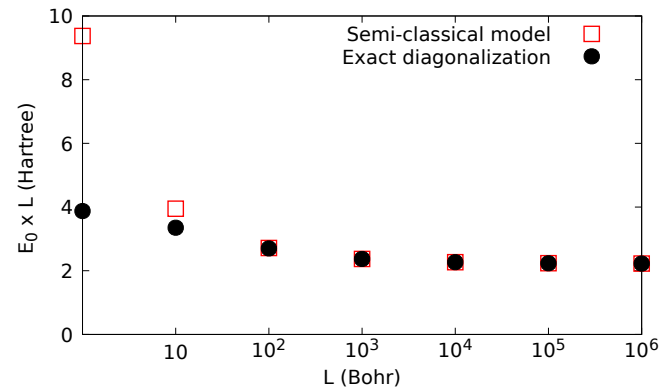}
    \caption{
     Scaled ground-state energies ($E_0 \times L$) for 2-dimensional CSC of various sizes. Open symbols: semi-classical model for the the low-density regime; filled symbols: exact diagonalization of the Hamiltonian.}
    \label{2D_Validation}
\end{figure}

\begin{figure}[h]
    \centering
    \includegraphics[width=\linewidth]{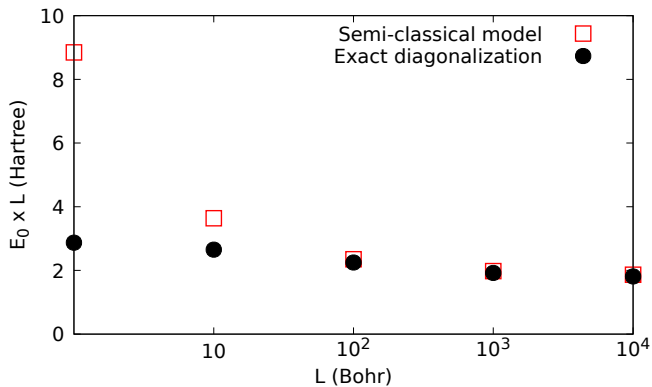}
    \caption{
     Scaled ground-state energies ($E_0 \times L$) for 3-dimensional CSC of various sizes.Open symbols: exact diagonalization of the Hamiltonian; filled symbols: semi-classical model for the the low-density regime.} 
    \label{3D_Validation}
\end{figure}
\begin{center}
\begin{figure}[!h]
    \includegraphics[width=\linewidth]{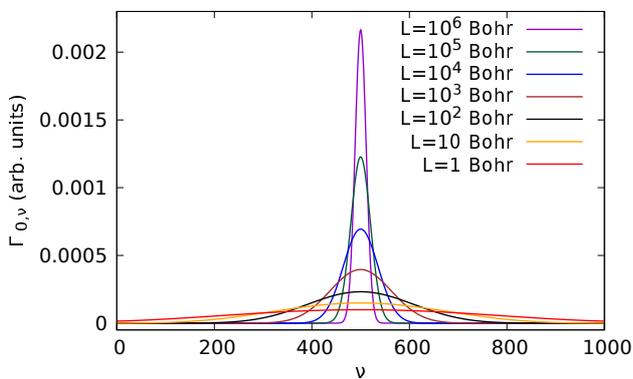}
\caption{$\Gamma_{0,\nu}$
as a function of $\nu$ for two electrons on a 1D Clifford torus
for various values of the system length $L$. 
The position of one electron is fixed around the center of the gaussian located at the origin ($\nu=0$).
Notice that the first $(\nu=0)$ and the last $(\nu=1000)$ points coincide.
} 
    \label{density_1D}%
\end{figure}
\end{center}
\begin{figure}[b] 
    \centering
    \includegraphics[trim=0cm 1cm 0cm 2cm, clip=true,width=\linewidth]{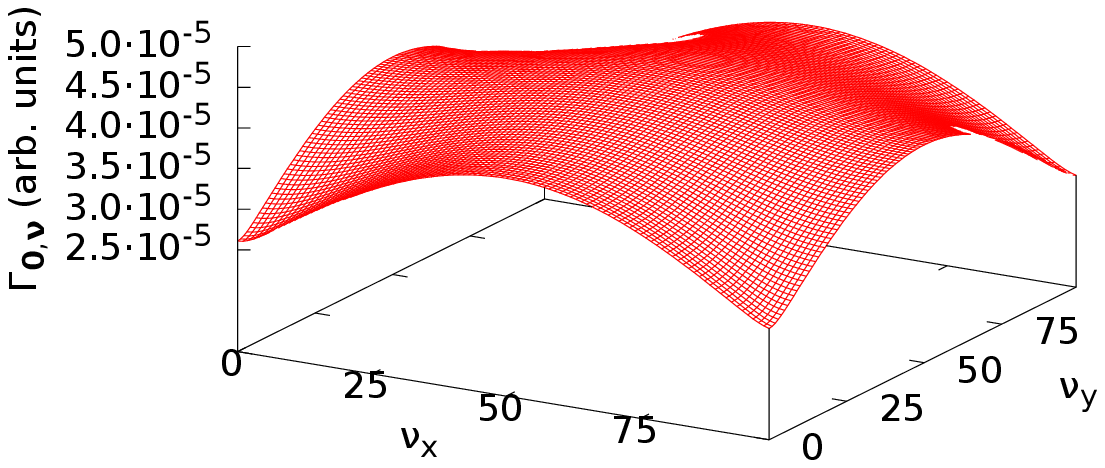} 
    \caption{ $\Gamma_{\mathbf{0},\boldsymbol{\nu}}$ 
    as a function of $\boldsymbol{\nu}$ for two electrons in a square 2-dimensional Clifford supercell with an edge of length $L=1$ bohr. 
    The position of one electron is fixed at the origin $\boldsymbol{\nu}=\mathbf{0}$.} 
\label{Density_2D_first}
\end{figure}
\begin{figure}[t]
    \centering
    \includegraphics[trim=0cm 1cm 0cm 2cm, clip=true, width=\linewidth]{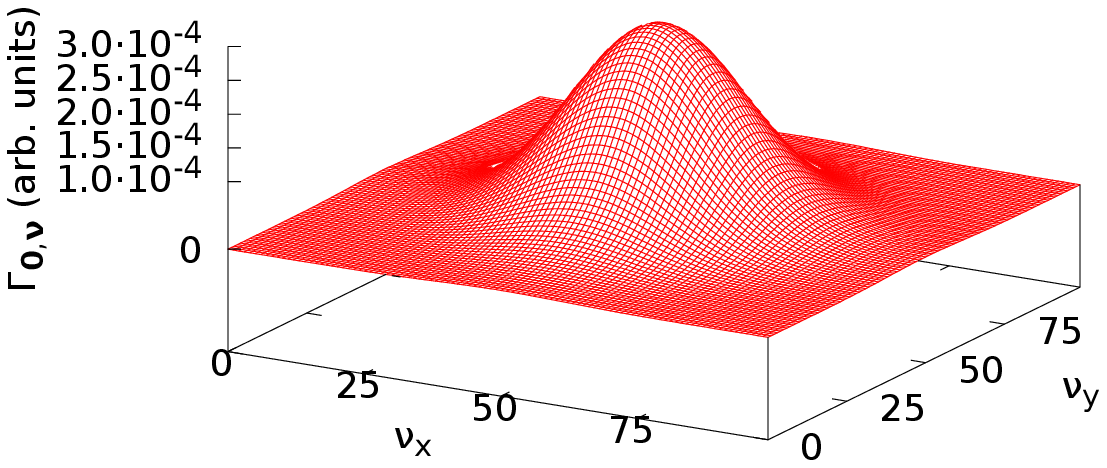} 
    \caption{$\Gamma_{\mathbf{0},\boldsymbol{\nu}}$ 
    as a function of $\boldsymbol{\nu}$ for two electrons in a square 2-dimensional Clifford supercell with an edge of length $L=100$ bohr. 
    The position of one electron is fixed at the origin $\boldsymbol{\nu}=\mathbf{0}$.} 
\end{figure}
\begin{figure}[t]
    \centering
    \includegraphics[trim=0cm 1cm 0cm 2cm, clip=true, width=\linewidth]{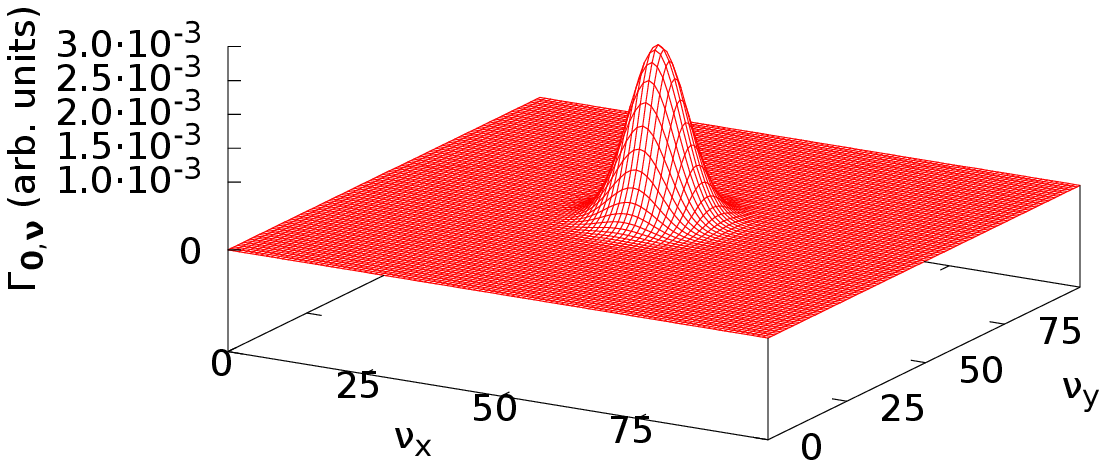} 
    \caption{$\Gamma_{\mathbf{0},\boldsymbol{\nu}}$ 
    as a function of $\boldsymbol{\nu}$ for two electrons in a square 2-dimensional Clifford supercell with an edge of length $L=10^4$ bohr. 
    The position of one electron is fixed at the origin $\boldsymbol{\nu}=\mathbf{0}$.} 
\end{figure}
\begin{figure}[t]
   \centering
\includegraphics[trim=0cm 1cm 0cm 2cm, clip=true, width=\linewidth]{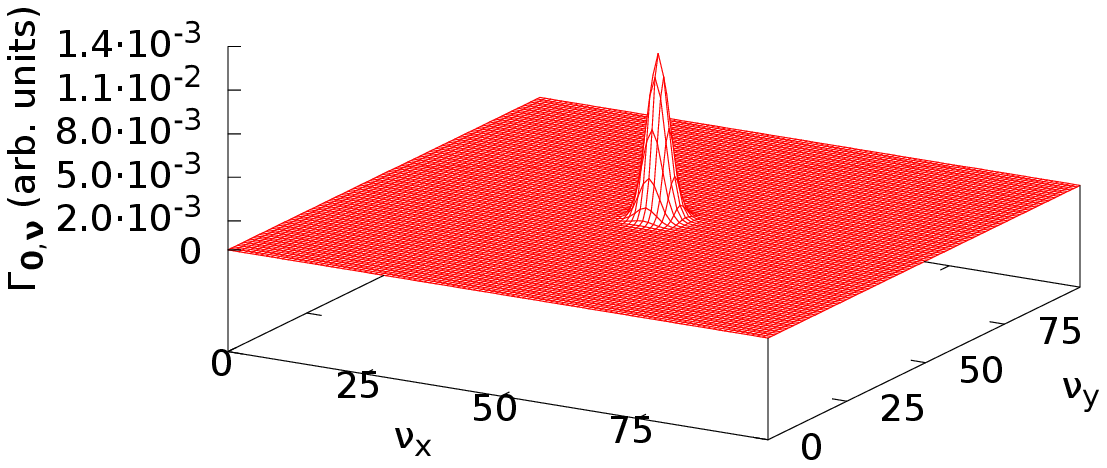} %
    \caption{$\Gamma_{\mathbf{0},\boldsymbol{\nu}}$ 
    as a function of $\boldsymbol{\nu}$ for two electrons in a square 2-dimensional Clifford supercell with an edge of length $L=10^6$ bohr. 
    The position of one electron is fixed at the origin $\boldsymbol{\nu}=\mathbf{0}$.} 
\label{Density_2D_last}
\end{figure}
\begin{figure}[b] 
    \centering
    \includegraphics[width=0.88\linewidth]{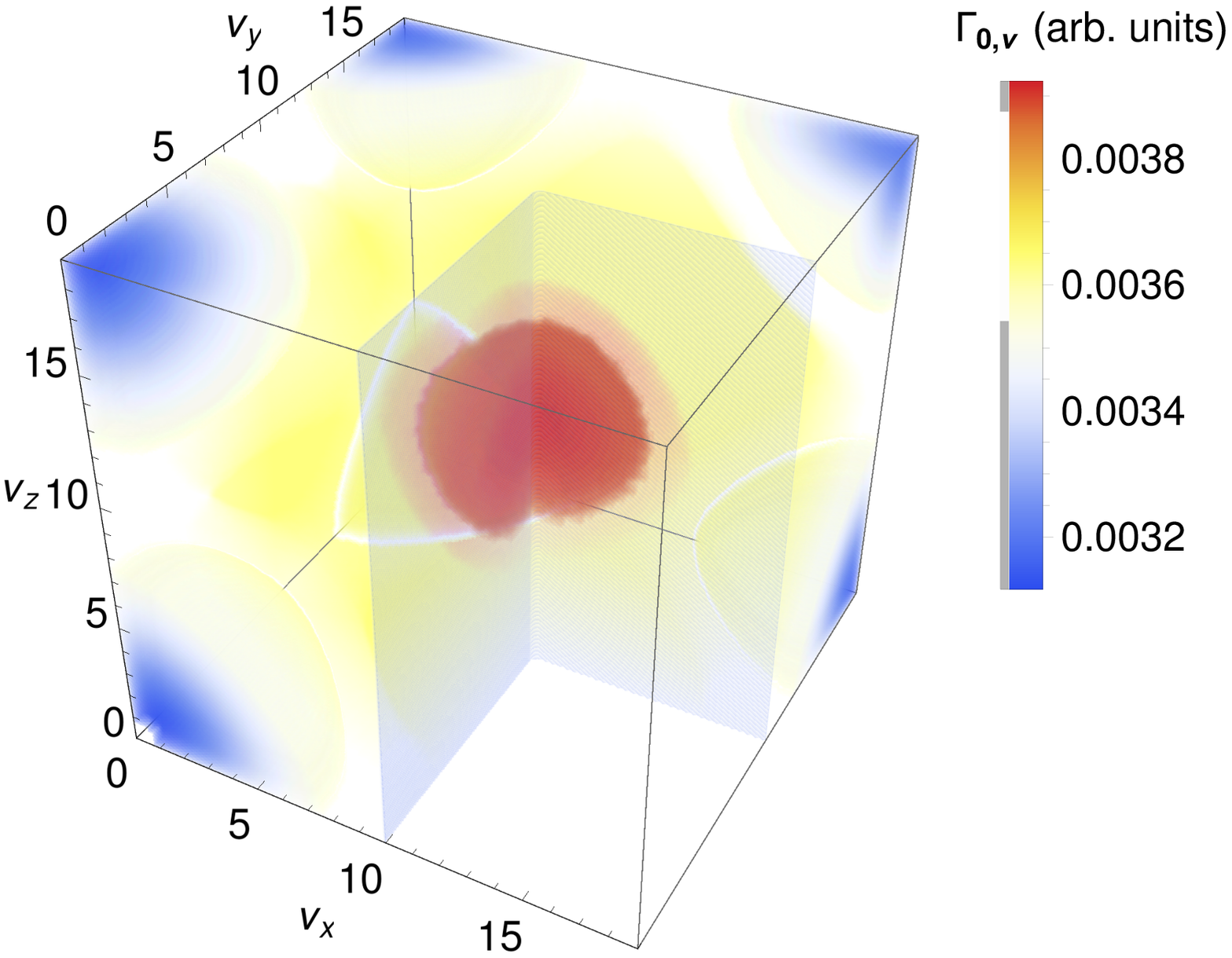} 
        \caption{ $\Gamma_{\mathbf{0},{\boldsymbol{\nu}}}$ 
    as a function of $\boldsymbol{\nu}$ for two electrons in a cubic 3-dimensional Clifford supercell with an edge of length $L=1$ bohr.
    The position of one electron is fixed at the origin $\boldsymbol{\nu}=\mathbf{0}$.} 
\label{Density_3D_first}
\end{figure}
\begin{figure}[t]
    \centering
    \includegraphics[width=0.88\linewidth]{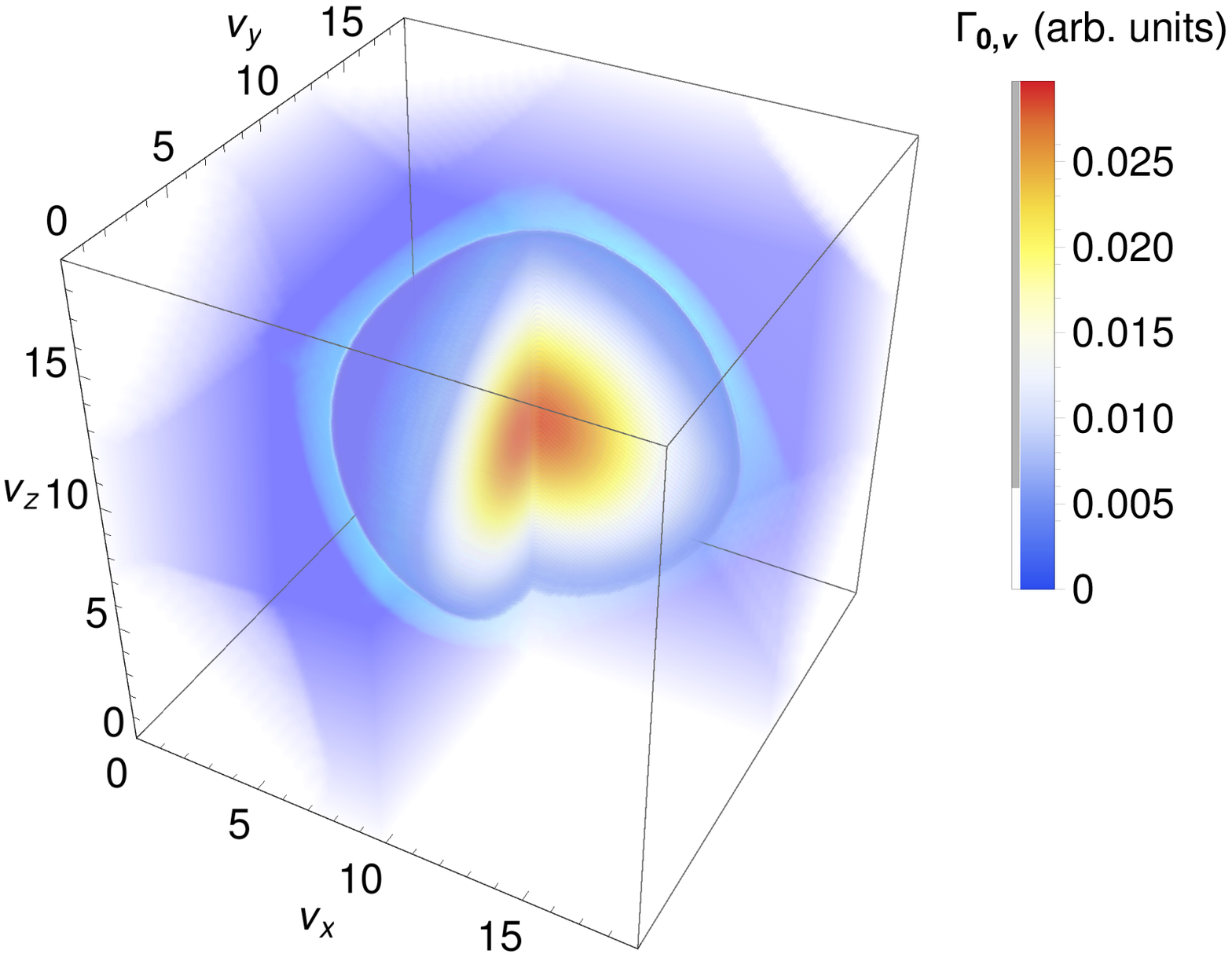}  
        \caption{ $\Gamma_{\mathbf{0},\boldsymbol{\nu}}$  
    as a function of $\boldsymbol{\nu}$ for two electrons in a cubic 3-dimensional Clifford supercell with an edge of length $L=100$ bohr.
    The position of one electron is fixed at the origin $\boldsymbol{\nu}=\mathbf{0}$.} 
\end{figure}
\begin{figure}[t] 
   \centering
\includegraphics[width=0.88\linewidth]{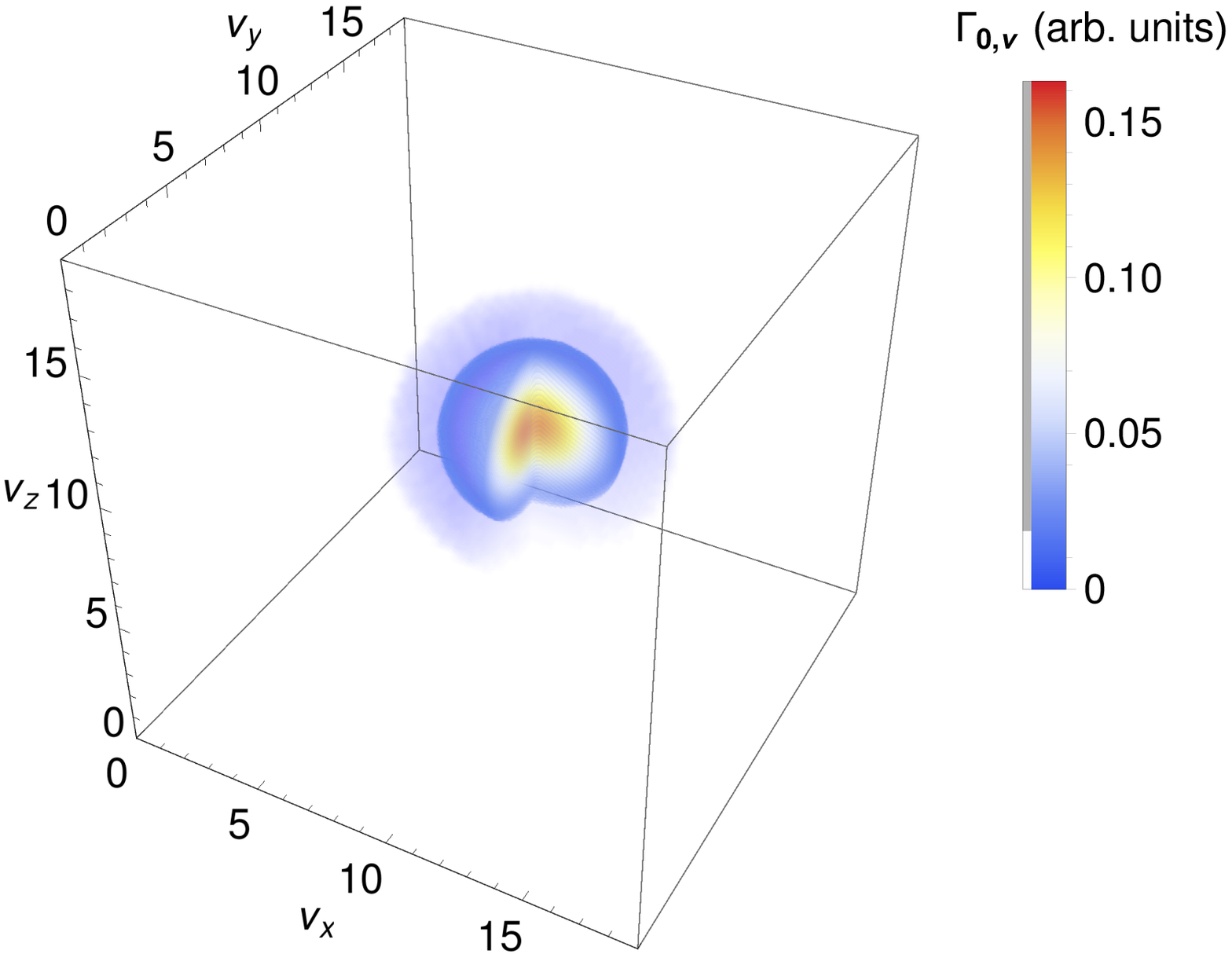} %
    \caption{$\Gamma_{\mathbf{0},\boldsymbol{\nu}}$ 
    as a function of $\boldsymbol{\nu}$ for two electrons in a cubic 3-dimensional Clifford supercell with an edge of length $L=10000$ bohr.
    The position of one electron is fixed at the origin $\boldsymbol{\nu}=\mathbf{0}$. } 
    \label{Density_3D_last}
\end{figure}
\begin{figure}[h]
    \centering
    \includegraphics[width=\linewidth]{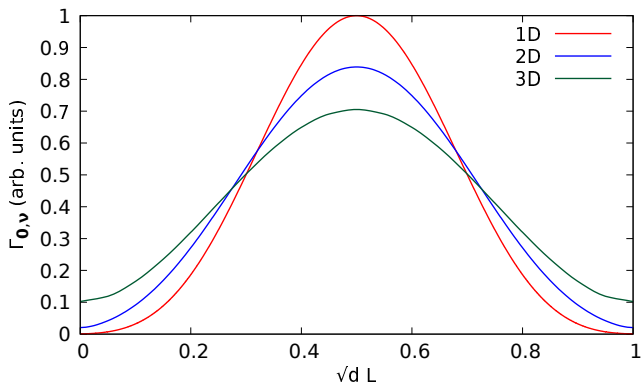}
\caption{The diagonal of $\Gamma_{\mathbf{0},\bnu}$ as a function of $\sqrt{d}L$ for systems of different dimensions with $L=10$ Bohr.
    The curves have been normalized such that the surface area underneath all three curves are equal.
    }
    \label{All_together}
\end{figure}
\begin{figure}[h] 
    \centering
    \includegraphics[width=\linewidth]{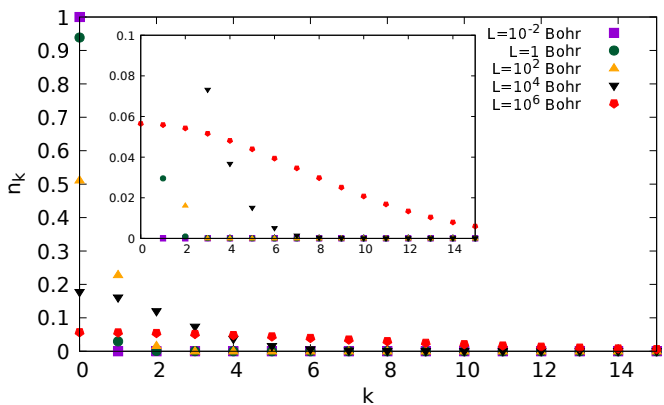}
        \caption{The natural occupation numbers $n_k$ as a function of $k$ for 1D Clifford tori of various sizes. We note that the occupation numbers of two spinorbitals having the same spatial part overlap.} 
\label{occupation}
\end{figure}
\begin{figure}[h] 
    \centering
    \includegraphics[width=\linewidth]{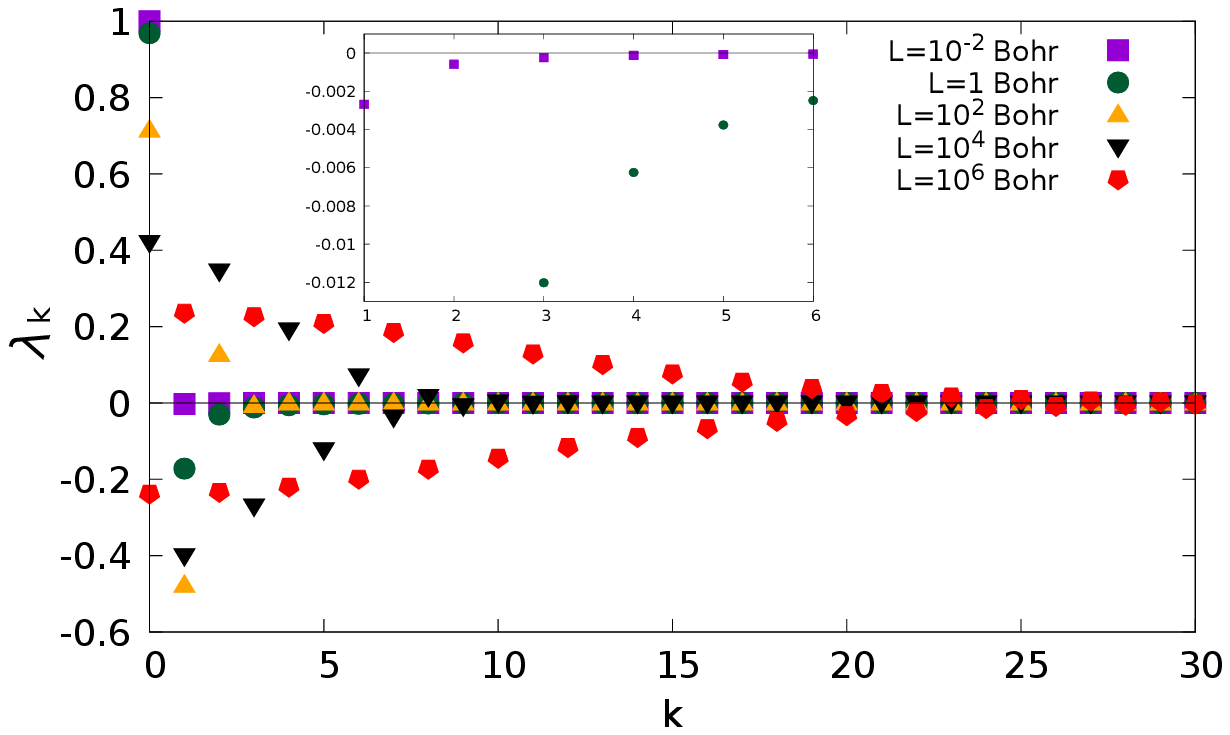}
        \caption{The natural amplitudes $\lambda_k$ as a function of $k$ for 1D Clifford tori of various sizes. We note that the amplitudes of two spinorbitals having the same spatial part overlap.} 
\label{amplitude}
\end{figure}

In 2D and 3D, only the ground-state energy obtained with our \emph{ab initio} approach
can be unambiguously identified with that of the semi-classical model but not the excited states because
there are many excited states that are degenerate in the model but not in the \emph{ab initio} calculations.
Therefore, in the following we will focus our comparison on the ground state only. 
In Figs.~\ref{2D_Validation} and \ref{3D_Validation} we collected 
the ground-state energies of systems of various sizes for 2- and 3- dimensional systems, respectively.
We observe a similar trend as for the 1D case, i.e., important differences between the model and the numerical calculations for small $L$ and an increasingly good agreement between the two methods for larger $L$.
We can conclude that our \emph{ab initio} approach can accurately describe the energies in the low-density regime in 1D, 2D and 3D.

Finally, we can also compare our results for the 2D Clifford torus to those obtained in Ref.~[\onlinecite{Loos_2009}]  in which the ground-state energies of two electrons confined on a sphere of radius $R$ are calculated. In the limit $R\rightarrow\infty$ the electrons will localize on opposite sides of the sphere and the ground-state energy will tend to $1/(2R)$. Since for the 2D Clifford torus the ground-state energy tends to $\pi/(\sqrt{2}L)$ in the limit $R\rightarrow\infty$ we expect that when $L=\sqrt{2}\pi R$ the ground-state energies of the two systems tend to the same value.
We have verified numerically that this is indeed the case.
We note that the conditions $L=\sqrt{2}\pi R$ implies that the sphere and the Clifford torus have different densities.

%
\subsection{Wigner localization}
As explained in section \ref{theory} the Wigner localization can be characterized by $\Gamma_{\boldsymbol{\mu},\boldsymbol{\nu}}$ which is the (diagonal of) the 2-RDM 
expressed in the basis of the gaussian orbitals because
it describes the probability of finding an electron contained in gaussian $g_{\boldsymbol{\mu}}$ 
when another electron is confined to gaussian $g_{\boldsymbol{\nu}}$.
In Fig.~\ref{density_1D} we report the ground-state $\Gamma_{0,\nu}$ as a function of $\nu$ for a one-dimensional CSC, i.e. one electron is kept fixed in the gaussian located at the origin.
We see that in the high-density region (small $L$) 
$\Gamma_{{0},{\nu}}$ is slowly varying. 
This is because in this regime the kinetic contribution to the energy dominates the electronic repulsion.
Therefore the electrons are delocalized and behave as in a free-electron gas.
Instead, in the low-density regime (large $L$), where the electronic repulsion dominates 
the kinetic energy, 
$\Gamma_{0,{\nu}}$ is peaked being non-negligible only in a small region of space.
This clearly indicates the Wigner localization of the electrons.
Not surprisingly, the position having the highest probability to find the second electron is at $L/2$ which
equals the largest possible distance between two electrons in 1D.
In Figs.~\ref{Density_2D_first}-\ref{Density_2D_last} we report $\Gamma_{\mathbf{0},\boldsymbol{\nu}}$ 
of a square 2-dimensional CSC
for $L=1$, $L=10^2$, $L=10^4$, and $L=10^6$ bohr, respectively.
As was the case in 1D, for small values of $L$ the 2D system behaves as a Fermi gas since
$\Gamma_{\mathbf{0},\boldsymbol{\nu}}$ 
is almost constant.
Instead, when the system size increases, we can clearly observe the electron localization from the peaked structure of $\Gamma_{\mathbf{0},\boldsymbol{\nu}}$ .
In 2D the position having the highest probability to find the second electron is in the middle of the square CSC, thus maximizing the distance between the two electrons.
In Figs.~\ref{Density_3D_first}-\ref{Density_3D_last} we report
$\Gamma_{\mathbf{0},\boldsymbol{\nu}}$ 
of a cubic 3-dimensional CSC for $L=1$, $L=10^2$, and $L=10^4$ bohr, respectively.
We note that $\Gamma_{\mathbf{0},\boldsymbol{\nu}}$ is represented by a color gradient, from blue (small) to red (large).
We ask the reader to pay attention to the fact that there are significant differences in the scale for the various figures.
Taking this into account we see once more that for small $L$ we have an almost constant 
$\Gamma_{\mathbf{0},\boldsymbol{\nu}}$ 
while for large $L$ we observe that 
$\Gamma_{\mathbf{0},\boldsymbol{\nu}}$ 
is localized in the middle of the cubic CSC, which is the furthest point
from the other electron located at the origin.

In order to compare the localization in the different dimensions we compare 
$\Gamma_{{0},{\nu}}$ corresponding to the one-dimensional CSC
with the diagonal elements of $\Gamma_{\mathbf{0},\boldsymbol{\nu}}$ for the two- and three-dimensial CSCs.
In Fig.~\ref{All_together} we report this comparison for $L=10$ bohr.
We observe that the amount of localization is proportional to the number of dimensions of the system, 
i.e., the localization is largest in 1D and smallest in 3D.
The reason is that, for fixed $L$, in higher dimensions the electrons have more space available to avoid each other.
We note that this result is consistent with those obtained for the transition of a Fermi liquid to a Wigner crystal in different dimensions. This  Wigner-Seitz radius at which this transition occurs is proportional to the number of dimensions~\cite{Ceperley_1980,Tanatar_1989,Rapisarda_1996,Drummond_2004,Drummond09}.
\subsection{Natural amplitudes and occupation numbers}

Although the one-body reduced density matrix (1-RDM) cannot be used to explicitly visualize the Wigner localization an implicit link can still be made. The eigenvalues of the 1-RDM are the natural occupation numbers and they can be related to the amount of electron correlation in the system~\cite{Giesbertz_2013,DiSabatino_2015}.
When all occupation numbers are either 0 or 1 the system can be considered uncorrelated and when all the occupation numbers have the same fractional value equal to the number of electrons divided by the number of spinorbitals the system can be considered maximally correlated~\cite{Escobar_2019}.
In the case of two electrons the natural occupation numbers $n_{\bk}$ are equal to $C_{\bk,-\bk}^2$, i.e., the square of the coefficients ($\bK=\mathbf{0}$) in Eq.~(\ref{psi_k})~\cite{Lowdin_1956}.
In Fig.~\ref{occupation} we report the natural occupation numbers $n_k$ for the 1D Clifford torus as a function of $k$.
As expected, we observe that for small $L$ (weak correlation) two natural spinorbitals, both having an identical spatial part, have an occupation close to 1 and all the other orbitals have an occupation close to 0, while for large $L$ (strong correlation) there are no natural orbitals with an occupation close to 1 and several orbitals with a small but finite occupation number.

It is also interesting to investigate the behavior of the natural amplitudes $\lambda_k$, which in our two-electron systems are equal to the coefficients $C_{\bk,-\bk}$.
We report these amplitudes as a function of $k$ for the 1D Clifford torus in Fig.~\ref{amplitude}.
We notice that for small $L$ (weak correlation) there is a single positive natural amplitude (at $k=0$), while for large $L$ (strong correlation) the positive and negative amplitudes alternate.
Similar trends have previously been observed for the two-electron harmonium atom, which is a model system consisting of two electrons that are confined by a harmonic potential.
For large harmonic frequencies (weak correlation) the harmonium atom has one positive natural amplitude while for small frequencies (strong correlation) there is an alternation of the positive and negative amplitudes~\cite{Cioslowski_2000}.
We defer the investigation of interesting related properties, such as the existence of solitonic natural orbitals and a possible power-law dependence of the decay of the occupation numbers as a function of $k$ to future studies~\cite{Cioslowski_2018,Cioslowski_2020}.

\section{Conclusions and Outlook}
\label{conclusions}

We have presented an accurate and numerically efficient approach to study Wigner localization in systems of various dimensions (1D, 2D, 3D). Its main features are: 1) the application of Clifford periodic boundary conditions with a renormalized distance to describe
the Coulomb potential and 2) the use of gaussian basis functions that are placed on a regular grid inside a Clifford supercell.
We have validated our method by comparing its results to those obtained within a semi-classical model that becomes
exact in the limit of vanishing density. Finally, using the two-body reduced density matrix, we have demonstrated that our approach can accurately capture the Wigner localization.

Our approach paves the way for several interesting future developments:
1) the generalization of our approach to more than 2 electrons in order to study Wigner crystals.
2) the implementation of our approach in Hartree-Fock (HF) theory and post-HF \emph{ab initio} approaches such as coupled-cluster theory.
3) the inclusion of ions in our approach which will allow the study of the solid state.

We will investigate these topics in future works.
\section{Acknowledgment}
J.A.B. thanks the French Agence Nationale de la Recherche (ANR) for financial support (grant agreement ANR-19-CE30-0011).
\section{Conflict of interest}
The authors have no conflicts to disclose.
\section{Data availability}
The data that support the findings of this study are available from the corresponding author upon reasonable request.
\appendix
\section{Translational symmetry of the symmetry adapted orbitals}
\label{App:SAO}
A gaussian basis function was defined in Eq.~\eqref{gaussian} which we repeat here for convenience.
\begin{equation} 
g_{\boldsymbol{\mu}}({\mathbf{r}- \mathbf{R}_{\boldsymbol{\mu}}})  =
\bigg(\frac{2\alpha}{\pi}\bigg)^\frac{3}{4}
e^{-\alpha |\mathbf{r}- \mathbf{R}_{\boldsymbol{\mu}}|^2},
\end{equation}
where $\bR_{\bmu}$ is the position of the center of the gaussian.
We consider $m^d$ equidistant gaussian functions with the nearest-neighbor distance equal to $\delta$ and we define $V=L^d$ the size of our system, where $L=m\delta$ with $m$ an integer.
We define 
\begin{align}
   \bM &= m\bn
   \label{Mdef}
   \\
   \bT&=L\bn=\delta \bM,
\end{align}
where
\begin{equation}
\bn = \sum_{i=1}^d n_i \be_i,
\end{equation}
in which $n_i$ are integers and $\be_i$ are unit vectors.
We have the following identity
\begin{equation}
g_{\boldsymbol{\mu}}(\br-\bR_{\bmu+\bM}+\bT) = g_{\boldsymbol{\mu}}(\br-\bR_{\boldsymbol{\mu}})
\label{id3}
\end{equation}
%
because $\bR_{\bmu+\bM} = \bR_{\bmu} + \delta\bM = \bR_{\mu} + \bT$.

The symmetry adapted orbitals were defined in Eq.~\eqref{Eqn:SAO}.
We repeat the definition here for convenience,
\begin{equation}
 \phi_{\mathbf{k}}(\mathbf{r})  = \frac{1}{m^{d/2}}
  \sum_{\boldsymbol{\mu}}
 e^{ i \frac{2\pi}{m} \mathbf{k} \cdot \boldsymbol{\mu}} 
 g(\mathbf{r}- \mathbf{R}_{\boldsymbol{\mu}}).
\end{equation}
They respect the translational symmetry, i.e.,
\begin{equation}
 \phi_{\mathbf{k}}(\br+\bT) = \phi_{\mathbf{k}}(\mathbf{r})
\end{equation}
{\em Proof}:
\begin{equation}
\phi_{\mathbf{k}}(\br+\bT)= \frac{1}{m^{d/2}}
\sum_{\boldsymbol{\mu}}
e^{ i \frac{2\pi}{m} \mathbf{k} \cdot \boldsymbol{\mu}} 
g(\mathbf{r} - \mathbf{R}_{\boldsymbol{\mu}}+\bT)
\end{equation}
Introducing the change of variable $\bnu = \bmu - \bM$ we obtain
\begin{align}
\phi_{\mathbf{k}}(\br+\bT)&= \frac{1}{m^{d/2}}
\sum_{\boldsymbol{\mu}}
e^{ i \frac{2\pi}{m} \mathbf{k} \cdot (\bnu+\bM)} 
g(\mathbf{r} - \mathbf{R}_{\bnu+\bM}+\bT)
 \\ &=
\frac{1}{m^{d/2}}
\sum_{\boldsymbol{\nu}}
e^{ i \frac{2\pi}{m} \mathbf{k} \cdot\bnu}
g(\mathbf{r} - \mathbf{R}_{\bnu}) = \phi_{\mathbf{k}}(\br) 
\end{align}
where we used Eq.~\eqref{id3} and the fact that $e^{ i \frac{2\pi}{m} \mathbf{k} \cdot \bM} = e^{ i 2\pi \mathbf{k} \cdot \bn} = 1$.
\section{One- and two-electron integrals in the symmetry adapted basis}
\label{App:integrals}
\subsection{One-electron integrals}
Because of the translational symmetry of the system and the equidistant basis functions
the matrix elements of the kinetic energy have the following properties,
\begin{align} 
 T_{\boldsymbol{\mu},\boldsymbol{\nu}}\; &= \; T_{\boldsymbol{\mu}+\boldsymbol{\delta},\boldsymbol{\nu}+\boldsymbol{\delta}} \label{prop1}\\
  T_{\boldsymbol{\mu},\boldsymbol{\nu}}\; &= \; T_{\mathbf{M}-\boldsymbol{\mu},\mathbf{M}-\boldsymbol{\nu}} \label{prop2}
\end{align}
where $\mathbf{M}$ is defined in Eq.~\eqref{Mdef}.
Let us now work out the one-electron integrals in the symmetry adapted basis.
We have the following expression in terms of the unnormalized SAO
\begin{align}
 \mathcal{T}_{\mathbf{k},\mathbf{k}'}  & = \langle \phi_{\mathbf{k} }
 |- \frac{1}{2}\nabla^2  | 
 \phi_{\mathbf{k}'} \rangle
\\ & =
\frac{1}{m^d}\sum_{\boldsymbol{\mu,\nu}}
e^{-i\frac{2\pi}{m}[\mathbf{k} \cdot {\boldsymbol{\mu}}-\mathbf{k}' \cdot {\boldsymbol{\nu}}]}
T_{\boldsymbol{\mu},\boldsymbol{\nu}}
\end{align}
where $T_{\boldsymbol{\mu},\boldsymbol{\nu}}$ are the 1-electron integrals in the gaussian basis defined in Eq.~\eqref{kin_AO}.
Making use of the property of $T_{\boldsymbol{\mu},\boldsymbol{\nu}}$ given in Eq.~\eqref{prop1} and setting
$\boldsymbol{\delta}=\boldsymbol{-\mu}$ we obtain
\begin{align}
 \mathcal{T}_{{\mathbf{k}},{\mathbf{k'}}}&=
 \frac{1}{m^d}
 \sum_{\boldsymbol{\mu,\nu}}
 e^{-i\frac{2\pi}{m}[\mathbf{k} \cdot {\boldsymbol{\mu}}-\mathbf{k}' \cdot {\boldsymbol{\nu}}]}  
   T_{\mathbf{0},\boldsymbol{\nu}-\boldsymbol{\mu}}
\\ & = 
 \frac{1}{m^d}
 \sum_{\boldsymbol{\mu,\nu}}
 e^{-i\frac{2\pi}{m}[(\mathbf{k}-\mathbf{k}') \cdot {\boldsymbol{\mu}}-\mathbf{k}' \cdot ({\boldsymbol{\nu}}-{\boldsymbol{\mu}})]}  
   T_{\mathbf{0},\boldsymbol{\nu}-\boldsymbol{\mu}}
\end{align}
Performing a change of variable ${\boldsymbol{\nu}'}={\boldsymbol{\nu}}-{\boldsymbol{\mu}}$ we can rewrite the above equation
according to
\begin{align}
\mathcal{T}_{{\mathbf{k}},{\mathbf{k'}}} &=
 \frac{1}{m^d}
 \sum_{\boldsymbol{\nu}}
 e^{-i\frac{2\pi}{m}(\mathbf{k}-\mathbf{k}') \cdot {\boldsymbol{\nu}}}
 \sum_{\boldsymbol{\nu'}}
 e^{i\frac{2\pi}{m}\mathbf{k} \cdot {\boldsymbol{\nu}'}}  
   T_{\mathbf{0},\boldsymbol{\nu}'}
\\ &=
    \delta_{\mathbf{\mathbf{k}}, \mathbf{\mathbf{k}'}} 
 \sum_{\boldsymbol{\nu}'} e^{i \frac{2\pi}{m}\mathbf{\mathbf{k}'} \cdot \boldsymbol{\nu}'} 
 T_{\mathbf{0},\boldsymbol{\nu}'}
\end{align}
The right-hand side of the above equation is complex but we can rewrite the above equation in terms of purely real quantities.
Using the property given in Eq.~\eqref{prop2} we obtain
\begin{equation}
\mathcal{T}_{{\mathbf{k}},{\mathbf{k'}}} =
\frac{\delta_{\mathbf{k}, \mathbf{k}'}}{2}
 \sum_{\boldsymbol{\nu'} } e^{i \frac{2\pi}{m} \mathbf{\mathbf{k}'} \cdot \boldsymbol{\nu}'}
[ T_{\mathbf{0},\boldsymbol{\nu}'} + 
T_{\mathbf{0},{\mathbf{M} - \boldsymbol{\nu}' } }]
\end{equation}
By splitting the right-hand side of the above expression into the sum of two terms and performing a change of variable $\boldsymbol{\mu}=\mathbf{M} - \boldsymbol{\nu}'$ in the second term 
we arrive at
\begin{align}
\mathcal{T}_{{\mathbf{k}},{\mathbf{k'}}} &=
\frac{\delta_{\mathbf{k}, \mathbf{k}'}}{2}
 \sum_{\boldsymbol{\mu} } 
\left[e^{i \frac{2\pi}{m} \mathbf{\mathbf{k}'} \cdot \boldsymbol{\mu}} + 
e^{i \frac{2\pi}{m} \mathbf{\mathbf{k}'} \cdot (\mathbf{M} - \boldsymbol{\mu})} 
\right]T_{\mathbf{0},\boldsymbol{\mu} }
\\ &=
\delta_{\mathbf{k}, \mathbf{k}'} 
 \sum_{\boldsymbol{\mu} }
 \cos \left( \frac{2\pi}{m} \mathbf{\mathbf{k}'} \cdot \boldsymbol{\mu}  \right)
T_{\mathbf{0},\boldsymbol{\mu}}
\end{align}
which is purely real.
In a similar way, we find that the overlap $\mathcal{S}_{{\mathbf{k}},{\mathbf{k'}}}$ between symmetry adapted orbitals is
\begin{equation}
    \mathcal{S}_{{\mathbf{k}},{\mathbf{k'}}} = \delta_{\mathbf{k}, \mathbf{k}'} 
 \sum_{\boldsymbol{\mu} }
 \cos \left( \frac{2\pi}{m} \mathbf{\mathbf{k}'} \cdot \boldsymbol{\mu}  \right)
S_{\mathbf{0},\boldsymbol{\mu}}
\end{equation}
where $S_{\mathbf{0},\boldsymbol{\mu}}=\langle g_{\mathbf{0}}|g_{\mathbf{\bmu}}\rangle$.
The expression in Eq.~\eqref{one-electron integral} is then obtained by accounting for the normalization of the SAO.
\subsection{Two-electron Integrals}
We have the following symmetry relations for the two-electron integrals,
\begin{align} 
 \langle \boldsymbol{\mu} \boldsymbol{\nu}|
 \boldsymbol{\rho} \boldsymbol{\sigma} \rangle &= \;
    \langle \boldsymbol{\mu}+\boldsymbol{\delta}, \boldsymbol{\nu}+\boldsymbol{\delta} 
    |\boldsymbol{\rho}+\boldsymbol{\delta}, \boldsymbol{\sigma}+\boldsymbol{\delta}\rangle
    \label{prop1'}\\
   \langle \boldsymbol{\mu} \boldsymbol{\nu}|\boldsymbol{\rho} \boldsymbol{\sigma} \rangle &= \;
    \langle \mathbf{M}- \boldsymbol{\mu}, \mathbf{M}-\boldsymbol{\nu}|\mathbf{M}-\boldsymbol{\rho}, \mathbf{M}-\boldsymbol{\sigma} \rangle \label{prop2'} 
\end{align}
We can now use a similar strategy as that used in the previous subsection for the two-electron integrals.
We obtain the following expression in terms of the unnormalized SAO
\begin{equation}
\langle \mathbf{k} \mathbf{k}' | \mathbf{k}'' \mathbf{k}'''\rangle_{\!u} = 
\frac{1}{m^{2d}}
 \sum_{\boldsymbol{\mu \nu \rho \sigma}} \!\!
 e^{-i \frac{2\pi}{m} \left[
 \mathbf{k} \cdot \boldsymbol{\mu} +
 \mathbf{k}'   \cdot \boldsymbol{\nu} -
 \mathbf{k}''  \cdot \boldsymbol{\rho} - 
 \mathbf{k}''' \cdot \boldsymbol{\sigma}
 \right]} 
 \langle \boldsymbol{\mu} \boldsymbol{\nu} | \boldsymbol{\rho} \boldsymbol{\sigma} \rangle
\end{equation}
%
Making use of the symmetry relation given in Eq.~\eqref{prop1'} and setting 
$\boldsymbol{\delta}$ as $-\boldsymbol{\mu}$ we can rewrite the above equation as
\begin{align} 
\label{67}
 \nonumber
\langle \mathbf{k} \mathbf{k}' | \mathbf{k}'' \mathbf{k}'''\rangle_{\!u} & = 
 \sum_{\boldsymbol{\mu \nu \rho \sigma}} \!\!
 e^{-i \frac{2\pi}{m} \left[
 \mathbf{k} \cdot \boldsymbol{\mu} +
 \mathbf{k}'   \cdot \boldsymbol{\nu} -
 \mathbf{k}''  \cdot \boldsymbol{\rho} - 
 \mathbf{k}''' \cdot \boldsymbol{\sigma}
 \right]} 
 \\ & \times
\langle \mathbf{0}, \boldsymbol{\nu} -\boldsymbol{\mu}
 | \boldsymbol{\rho} -\boldsymbol{\mu}, \boldsymbol{\sigma} -\boldsymbol{\mu} \rangle
 \\ &=
  \nonumber
 \frac{1}{m^{2d}}
 \sum_{\boldsymbol{\mu}}
 e^{-i \frac{2\pi}{m}
 \left[\mathbf{k}+\mathbf{k}' -\mathbf{k}'' -\mathbf{k}''') \cdot \boldsymbol{\mu} \right]}
 \\ & \times
  \nonumber
 \sum_{\boldsymbol{\nu \rho \sigma}}
 e^{-i \frac{2\pi}{m} \big( \mathbf{k}' \cdot (\boldsymbol{\nu} - \boldsymbol{\mu}) -
 \mathbf{k}''  \cdot (\boldsymbol{\rho} - \boldsymbol{\mu}) -
 \mathbf{k}''  \cdot (\boldsymbol{\sigma} - \boldsymbol{\mu})
 \big)}
 \\ &\times
 \langle \mathbf{0}, \boldsymbol{\nu} -\boldsymbol{\mu}
 | \boldsymbol{\rho} -\boldsymbol{\mu}, \boldsymbol{\sigma} -\boldsymbol{\mu} \rangle
\end{align}
Performing the following changes of variables $\boldsymbol{\nu}' = \boldsymbol{\nu} -\boldsymbol{\mu}$,
$\boldsymbol{\rho}' = \boldsymbol{\rho} -\boldsymbol{\mu}$ and $\boldsymbol{\sigma}' = \boldsymbol{\sigma} -\boldsymbol{\mu}$
we arrive at
\begin{align} 
 \nonumber
\langle \mathbf{k} \mathbf{k}' | \mathbf{k}'' \mathbf{k}'''\rangle_{\!u} & = 
 \frac{1}{m^{2d}}
 \sum_{\boldsymbol{\mu}}
 e^{-i \frac{2\pi}{m}
 \left[\mathbf{k}+\mathbf{k}' -\mathbf{k}'' -\mathbf{k}'''\right]\cdot \boldsymbol{\mu}}
 \\ & \times
  \nonumber
 \sum_{\boldsymbol{\nu' \rho' \sigma'}}
 e^{-i \frac{2\pi}{m} \left( \mathbf{k}' \cdot \boldsymbol{\nu}' -
 \mathbf{k}''  \cdot \boldsymbol{\rho}' -
 \mathbf{k}''  \cdot \boldsymbol{\sigma}'
 \right)}
 \\ &\times
 \langle \mathbf{0}, \boldsymbol{\nu}'
 | \boldsymbol{\rho}', \boldsymbol{\sigma}' \rangle
 \\ &=
  \nonumber
  \frac{1}{m^d}\delta_{\mathbf{k}+\mathbf{k}' -\mathbf{k}'' -\mathbf{k}''}
 \\ & \times
 \sum_{\boldsymbol{\nu' \rho' \sigma'}}
 e^{-i \frac{2\pi}{m} \left( \mathbf{k}' \cdot \boldsymbol{\nu}' -
 \mathbf{k}''  \cdot \boldsymbol{\rho}' -
 \mathbf{k}''  \cdot \boldsymbol{\sigma}'
 \right)}
 \langle \mathbf{0}, \boldsymbol{\nu}'
 | \boldsymbol{\rho}', \boldsymbol{\sigma}' \rangle
\end{align}
Using Eq.~\eqref{prop2'} we can now rewrite the above expression in terms of purely real quantities.
We obtain
\begin{align} 
 \nonumber
\langle \mathbf{k} \mathbf{k}' | \mathbf{k}'' \mathbf{k}'''\rangle_{\!u} & =
  \nonumber
  \frac{1}{2m^d}\delta_{\mathbf{k}+\mathbf{k}' -\mathbf{k}'' -\mathbf{k}''}
 \\ & \times
    \nonumber
 \sum_{\boldsymbol{\nu' \rho' \sigma'}}
 e^{-i \frac{2\pi}{m} \left( \mathbf{k}' \cdot \boldsymbol{\nu}' -
 \mathbf{k}''  \cdot \boldsymbol{\rho}' -
 \mathbf{k}''  \cdot \boldsymbol{\sigma}'
 \right)}
 \\ &\times
 \left[
 \langle \mathbf{0}, \boldsymbol{\nu}'
 | \boldsymbol{\rho}', \boldsymbol{\sigma}' \rangle +
 \langle \mathbf{0}, \mathbf{M} -  \boldsymbol{\nu}'
 | \mathbf{M} - \boldsymbol{\rho}', \mathbf{M} - \boldsymbol{\sigma}' \rangle
 \right]
\end{align}
Splitting the right-hand side of the above expression into the sum of two terms and performing the changes of variable $\boldsymbol{\nu}= \mathbf{M}- \boldsymbol{\nu}'$, $
\boldsymbol{\rho}= \mathbf{M}- \boldsymbol{\rho}'$ and 
$\boldsymbol{\sigma}= \mathbf{M}- \boldsymbol{\sigma}'$
we arrive at the final expression for the 2-electron integrals expressed in the symmetry-adapted basis,
\begin{align} 
 \nonumber
\langle \mathbf{k} \mathbf{k}' | \mathbf{k}'' \mathbf{k}'''\rangle_{\!u} & =
  \nonumber
  \frac{1}{2 m^d}\delta_{\mathbf{k}+\mathbf{k}' -\mathbf{k}'' -\mathbf{k}''}
 \\ & \times
    \nonumber
 \sum_{\boldsymbol{\nu \rho \sigma}}
 \Big[
 e^{-i \frac{2\pi}{m} \left( \mathbf{k}' \cdot \boldsymbol{\nu}' -
 \mathbf{k}''  \cdot \boldsymbol{\rho}' -
 \mathbf{k}''  \cdot \boldsymbol{\sigma}'
 \right)} 
 \\ &+
  e^{-i \frac{2\pi}{m} \left( \mathbf{k}' \cdot (\mathbf{M} - \boldsymbol{\nu}) -
 \mathbf{k}''  \cdot (\mathbf{M} - \boldsymbol{\rho}) -
 \mathbf{k}''  \cdot (\mathbf{M} - \boldsymbol{\sigma})
 \right)}
 \Big]
\nonumber \\ &\times
 \langle \mathbf{0}, \boldsymbol{\nu}
 | \boldsymbol{\rho}, \boldsymbol{\sigma} \rangle
 \\ &=
   \nonumber
  \frac{1}{m^d}\delta_{\mathbf{k}+\mathbf{k}' -\mathbf{k}'' -\mathbf{k}''}
 \\ & \times
    \nonumber
 \sum_{\boldsymbol{\nu \rho \sigma}}
 \cos\left[\frac{2\pi}{m} \left( \mathbf{k}' \cdot \boldsymbol{\nu} -
 \mathbf{k}''  \cdot \boldsymbol{\rho} -
 \mathbf{k}''  \cdot \boldsymbol{\sigma}\right)\right]
 \\ &\times
 \langle \mathbf{0}, \boldsymbol{\nu}
 | \boldsymbol{\rho}, \boldsymbol{\sigma} \rangle
\end{align}
which is purely real.
We arrive at the expression in Eq.~\eqref{two-electron integral} by accounting for the normalization of the SAO.

\end{document}